\documentstyle[preprint,aps]{revtex}

\begin{document}

\draft

%\preprint{xxx-xx/95}

\title{Classical and Quantum Anisotropic Wormholes in Pure \\
General Relativity}

\author{Hongsu Kim}

\address{Department of Physics\\
Ewha Women's University, Seoul 120-750, KOREA \\
and \\
Department of Astronomy and Atmospheric Sciences \\
Kyungpook National University, Taegu, 702-701, KOREA \footnote{present address,
e-mail : hongsu@vega.kyungpook.ac.kr}}

\date{September, 1998}

\maketitle

\begin{abstract}
In the homogeneous and isotropic Friedmann-Robertson-Walker minisuperspace model,
it is known that there are no Euclidean wormhole solutions in the pure gravity
system. Here it is demonstrated explicitly that in Taub cosmology, which is one
of the simplest anisotropic cosmology models, wormhole solutions do exist in 
pure general relativity in both classical and quantum contexts.
\end{abstract}


\narrowtext
%\twocolumn

 Some time ago, the ``Euclidean wormhole physics'' stimulated an enormous 
excitement in the theoretical physics society due to the possible, provocative
effects it may have on low-energy physics. For instance, the wormhole
configuration may result in the generic loss of quantum coherence [2,3] or
provide a mechanism for fixing the observable value of the cosmological
constant to zero [1,2]. As for the quantitative aspect of this wormhole
physics, the construction of explicit wormhole instanton solutions had been
actively attempted [4-6]. To our disappointment, however, in the simplest
minisuperspace model, homogeneous and isotropic wormhole solutions had
been found only in the presence of somewhat exotic matter fields [8] but not
in the pure gravity system. In this letter, we shall demonstrate that 
wormholes with the simplest anisotropic structure do exist in pure
general relativity in both classical and quantum contexts. Classically,
wormholes are Euclidean metrics which are solutions to the Euclidean
classical field equations representing spacetimes consisting of two
asymptotically Euclidean regions joined by a narrow tube. In the quantum
sense, on the other hand, and particularly in the context of canonical
quantum cosmology, ``quantum wormholes'' [8] may be identified with a state 
represented by a solution to the Wheeler-DeWitt (WD) equation satisfying
a certain boundary condition [7] describing the typical wormhole
configuration.
Now consider the system of pure Einstein gravity with ``negative''
cosmological constant, i.e., the anti-de Sitter spacetime in 4-dim.
Its Einstein-Hilbert action and the associated Einstein field equation
are given respectively by
\begin{eqnarray}
S = {1\over 16\pi G} &[& \int_{M}d^4x\sqrt{g}(R-2\Lambda) + 2\int_{\partial M}
d^3x\sqrt{h}(K-K_{0})], \nonumber \\
R_{\mu\nu} &-& {1\over 2}g_{\mu\nu}R + \Lambda g_{\mu\nu} = 0 
\end{eqnarray}
where we introduced the Gibbons-Hawking surface term [9] on the boundary
$\partial M$ and $\Lambda = - |\Lambda |<0$.
We stress that the addition of negative cosmological constant here is 
not the necessary condition for the existence of wormhole solutions as we
shall see in a moment. Then, in order to look for ``anisotropic''
(both classical and quantum) wormhole solutions, we choose to work in
``Taub cosmology model'' [10] (which is one of the simplest homogeneous,
anisotropic cosmology models known) described by the metric
\begin{eqnarray}
ds^2 = - N^2(t)dt^2 + e^{2\alpha(t)} (e^{2\beta(t)})_{ij} \sigma^{i}
\otimes \sigma^{j}
\end{eqnarray} 
where the Euclidean signature metric can be obtained by replacing $t$ by
the Euclidean time $\tau = it$ and thus $- N^2(t)dt^2$ by 
$N^2(\tau)d\tau^2$. $N(t)$ and $a(t)=e^{\alpha(t)}$ denote the lapse
function and the scale factor respectively and $\beta_{ij}(t) = 
diag(\beta, \beta, -2\beta)$ represents the degree of anisotropy. Thus 
this Taub model can be thought of as a special case of more general 
Bianchi type-IX model of anisotropic cosmology. Namely in the Bianchi
type-IX model, the anisotropy measure $\beta_{ij}$ has two independent
components $\beta_{+}$ and $\beta_{-}$. And the special case in which
one sets $\beta_{-}=0$ keeping $\beta_{+} \equiv \beta \neq 0$ amounts to
the reduction of the model to the Taub model. The non-holonomic basis
$\{\sigma^{i}\}$ form a basis of 1-forms on a spacelike 3-sphere $S^3$
satisfying the SU(2) Maurer-Cartan structure equation
\begin{eqnarray}
d\sigma^{i} = {1\over 2}\epsilon^{ijk}\sigma^{j}\wedge \sigma^{k}
\end{eqnarray}
and can be represented in terms of 3-Euler angles $0\leq \theta \leq \pi$,
$0\leq \phi \leq 2\pi$, $0\leq \psi \leq 4\pi$ parametrizing $S^3$ as
\begin{eqnarray}
\sigma^{1} &=& \cos \psi d\theta + \sin \psi \sin \theta d\phi, \nonumber \\
\sigma^{2} &=& \sin \psi d\theta - \cos \psi \sin \theta d\phi, \\
\sigma^{3} &=& d\psi + \cos \theta d\phi. \nonumber
\end{eqnarray}
The two independent metric functions, $a(t)=e^{\alpha(t)}$ and $\beta(t)$
thus can be regarded as two ``minisuperspace variables'' $\gamma^{A} = 
(a, \beta)$ and in terms of them the gravity action (dropping the
Gibbons-Hawking surface term) can be written as 
\begin{eqnarray}
S = {1\over 2}\int dt Na^3 [{1\over N^2}\{-({\dot{a}\over a})^2+\dot{\beta}^2\}
- \{{1\over a^2}V(\beta) + {4\over 3}\Lambda \}]
\end{eqnarray}
where we defined the ``curvature potential'' $V(\beta)$ as
\begin{eqnarray}
V(\beta) \equiv {1\over 3}(e^{-8\beta} - 4 e^{-2\beta})
\end{eqnarray}
and we redefined the action by multiplying it by an overall constant,
$S\rightarrow (3\pi/4G)S$ and then rescaled the lapse function as
$N\rightarrow 2N$. For later use, we also provide, in the Lorentzian
signature, the expressions for the curvature scalar, $R$ of the 4-dim.
space and the trace of the extrinsic curvature, $K$ of a 3-dim.
spacelike hypersurface ;
\begin{eqnarray}
R &=& {6\over N^2}[({\ddot{a}\over a}) + ({\dot{a}\over a})^2 + \dot{\beta}^2
- ({\dot{N}\over N})({\dot{a}\over a})] - {3\over 2a^2}V(\beta), \nonumber \\
K &=& -{3\over N}({\dot{a}\over a}).
\end{eqnarray}
Next, upon identifying the momenta $\pi_{A} = (P_{a},
P_{\beta})$ conjugate to the minisuperspace variables $\gamma^{A} =
(a, \beta)$ as $P_{a}=\partial S/\partial \dot{a} = -a\dot{a}/N$ and
$P_{\beta}=\partial S/\partial \dot{\beta} = a^3\dot{\beta}/N$, one can go
over to the Hamiltonian of the system via the Legendre transformation as
\begin{eqnarray}
S &=& \int dt L_{ADM} = \int dt (P_{a}\dot{a}+P_{\beta}\dot{\beta}-H_{ADM})
\nonumber \\
{\rm with} \qquad H_{ADM} &=& NH_{0} + N_{i}H^{i}, \qquad N_{i}=0, \\
H_{0} &=& -{\delta S\over \delta N} = {1\over 2}a^{-3}[-a^2P^2_{a}+P^2_{\beta}
+a^4\{V(\beta) + {4\over 3}\Lambda a^2\}] \nonumber
\end{eqnarray}
which (i.e., $H_{0}$) vanishes identically due to the time-reparametrization
invariance in general relativity. This is called the classical Hamiltonian
constraint, $H_{0} = 0$ and, upon Dirac quantization, it turns into the
Wheeler-DeWitt equation to which we shall come back later on. 
In order to find classical wormhole solutions, which are Euclidean objects,
from this point on, we need to work in Euclidean signature. The Euclidean 
gravity action can be obtained, via the analytic continuation, as
\begin{eqnarray}
I = -iS = {1\over 2}\int d\tau Na^3 [{1\over N^2}\{-({a'\over a})^2 + \beta'^2\}
+ \{{1\over a^2}V(\beta) + {4\over 3}\Lambda\}]
\end{eqnarray}
where now the ``prime'' denotes the derivative with respect to the Euclidean
time $\tau$ while earlier the ``overdot'' denoted that with respect to the
Lorentzian time $t$. By varying this Euclidean action with respect to the
lapse $N(\tau)$, once again one gets the Hamiltonian constraint
\begin{eqnarray}
{1\over N^2}({da\over d\tau})^2 - {a^2\over N^2}({d\beta\over d\tau})^2 +
[V(\beta) + {4\over 3}\Lambda a^2] = 0
\end{eqnarray}
where $(d\beta/d\tau) = -i(d\beta/dt) = -i(N/a^3)P_{\beta}$ using the Wick
rotation rule, $\tau = it$ while varying it with respect to minisuperspace
variables $a, \beta$, one gets, respectively,
\begin{eqnarray}
{1\over N^2a}({d^2a\over d\tau^2}) &+& {2\over N^2}({d\beta \over d\tau})^2 +
{4\over 3}\Lambda = 0, \\
{1\over N^2}({d^2\beta \over d\tau^2}) &+& {3\over N^2a}({da\over d\tau})
({d\beta \over d\tau}) - {1\over 2a^2}({dV\over d\beta}) = 0.
\end{eqnarray}
Note, here, that the Hamiltonian constraint in eq.(10) coincides with the
$\tau \tau$-component of the Einstein equation, 
$R_{\tau \tau}-{1\over 2}g_{\tau \tau}R+\Lambda g_{\tau \tau}=0$ and to
arrive at the expression for the eq.(11), we made use of the Hamiltonian
constraint in eq.(10). Needless to say, constructing a general solution
of the coupled equations (10),(11) and (12) is a highly non-trivial job.
However, by restricting our interest to some particular class of wormhole
solutions with specific character, we can have some insight into the
existence of classical wormhole solutions. Namely, suppose we particularly 
look for wormhole solutions with very small/large anisotropy,
\begin{eqnarray}
V(\beta) &=& {1\over 3}(e^{-8\beta} - 4 e^{-2\beta}) \rightarrow k =
-1 (\beta\rightarrow 0), ~~0 (\beta\rightarrow \infty), \\
{dV\over d\beta} &=& {8\over 3}(-e^{-8\beta} + e^{-2\beta}) \rightarrow 0
(\beta\rightarrow 0 ~~{\rm or} ~~~\beta\rightarrow \infty)
\end{eqnarray}
and with ``constant'' momentum associated with the anisotropy change,
\begin{eqnarray}
P_{\beta} = i {a^3\over N}\beta' = const.
\end{eqnarray}
Imposing these conditions for the character of the solution, the eqs.(10),
(11) and (12) above reduce to, respectively,
\begin{eqnarray}
{1\over N^2}({da\over d\tau})^2 + {P^2_{\beta}\over a^4} + [k + {4\over 3}
\Lambda a^2] = 0, \\
{1\over N^2a}({d^2a\over d\tau^2}) - {2\over a^6}P^2_{\beta} + {4\over 3}
\Lambda = 0, \\
{1\over N^2}{d\over d\tau}({N\over a^3}) + {3\over Na^4}({da\over d\tau})
= 0. 
\end{eqnarray}
Next, we now fix the gauge associated with the time-reparametrization
invariance as $N(\tau) = 1$. Incidentally, then, the eq.(18)
above is automatically satisfied and we are left just with the 
eq.(16) which is an ODE for the scale factor $a(\tau)$ alone.
(Eqs.(16) and (17), upon fixing the gauge $N(\tau) = 1$,
become the same since one can get the latter by differentiating the former
with respect to the Euclidean time $\tau$.) Thus what remains is to solve
the eq.(16), i.e.,
\begin{eqnarray}
({da\over d\tau})^2 + [k + {4\over 3}\Lambda a^2] = -{P^2_{\beta}\over a^4}
\end{eqnarray}
for $a(\tau)$ and then plug it into 
\begin{eqnarray}
\beta(\tau) = -iP_{\beta}\int^{\tau}_{-\infty}{d\tau'\over a^3(\tau')}
\end{eqnarray}
to obtain the behavior of anisotropy. Here, we set the ``initial'' time to
be $\tau \rightarrow -\infty$ when yet neither the wormholes nor the baby
universes were born and no anisotropy yet arises, i.e., $\beta(-\infty) = 0$.
At this point, it seems worth noting
that after fixing the gauge $N(\tau)=1$, it becomes clear that the condition
given in eq.(15), i.e., the constant momentum associated with the change
in anisotropy, $P_{\beta} = const.$ corresponds to the condition for
wormhole solutions whose anisotropy changing rate is inversely proportional
to the scale (i.e., the size) of the wormhole, $(d\beta/d\tau) = (-i)
P_{\beta}/a^3$, which seems quite a reasonable expectation. We now attempt
to solve the eq.(19) for exact wormhole solutions. \\
(i) wormhole solution with very small anisotropy ($\beta\rightarrow 0$) \\
This case amounts to choosing $k = -1$, and then the eq.(19) admits
an exact solution in the absence of the cosmological constant. Namely
$(da/d\tau)^2 - 1 = - P^2_{\beta}/a^4$  yields, upon integration,
\begin{eqnarray}
\tau &=& \int^{\tau}_{0} d\tau' =
\int^{a(\tau)}_{a(0)} {a^2da\over \sqrt{(a^2+P_{\beta})(a^2-P_{\beta})}}\\
&=& \sqrt{{P_{\beta}\over 2}}F[\arccos ({\sqrt{P_{\beta}}\over a}),
{1\over \sqrt{2}}] - \sqrt{2P_{\beta}}E[\arccos ({\sqrt{P_{\beta}}\over a}),
{1\over \sqrt{2}}] + {1\over a}\sqrt{a^4 - P^2_{\beta}} \nonumber
\end{eqnarray}
where we chose that the wormhole throat, i.e., the minimum value of
$a(\tau)$ occurs for $\tau=0$, namely $({da\over d\tau})|_{\tau=0}=0$
and $a(\tau=0)=\sqrt{P_{\beta}}$ and $F$ and $E$ are the elliptic integrals
of the 1st and 2nd kind respectively. This solution can be identified
with a classical wormhole configuration since it displays the
asymptotic behavior $a^2(\tau)\rightarrow \tau^2$ as $\tau\rightarrow
\pm \infty$. \\
Now that we have found the exact wormhole solution in closed form.
Then next, we discuss precisely what topology changing process this wormhole
instanton solution describes. To this end, we note ; \\
For $\tau \rightarrow -\infty$,
\begin{eqnarray}
&a^2(\tau) \rightarrow \tau^2 \rightarrow \infty, ~~~~({da\over d\tau})|_{\tau
\rightarrow -\infty} \rightarrow 1, \nonumber \\
&\beta(-\infty) = 0, ~~~~({d\beta \over d\tau})|_{\tau \rightarrow -\infty} =
-iP_{\beta}{1\over a^3(-\infty)} \rightarrow 0 \nonumber
\end{eqnarray}
thus the curvature scalar there goes like $R(\tau \rightarrow -\infty)
\rightarrow 0$, namely $\tau \rightarrow -\infty$ spacelike hypersurface
is flat. \\
For $\tau = 0$,
\begin{eqnarray}
&a(0) = \sqrt{P_{\beta}}, ~~~~({da\over d\tau})|_{\tau = 0} = 0, \nonumber \\
&\beta(0) = -iP_{\beta}\int^{0}_{-\infty}{d\tau \over a^3(\tau)} = const., 
~~~~({d\beta \over d\tau})|_{\tau = 0} = -iP_{\beta}{1\over a^3(0)} = -i
{1\over \sqrt{P_{\beta}}} \nonumber
\end{eqnarray}
thus the trace of the extrinsic curvature there is $K(\tau = 0) = 3({a'\over a})|_{\tau
= 0} = 0$, namely this $\tau = 0$ spacelike hypersurface is a surface of vanishing
extrinsic curvature and corresponds to the wormhole throat. And this surface has
small, constant anisotropy and constant anisotropy changing rate.
To conclude, this wormhole solution with very small anisotropy is an instanton
configuration representing the topology change from the initial ($\tau \rightarrow -\infty$)
state with topology $R^3$ to the final ($\tau = 0$) state with the topology $R^3 \oplus S^3$,
i.e., flat space with an additional closed baby universe. \\
(ii) wormhole solution with very large anisotropy ($\beta\rightarrow \infty$)\\
This case amounts to choosing $k=0$ and the eq.(19) admits an exact
solution even in the presence of the cosmological constant. Namely
$(da/d\tau)^2 - 4|\Lambda|a^2/3 = - P^2_{\beta}/a^4$, upon integration,
\begin{eqnarray}
\tau = \int^{\tau}_{0} d\tau' =
\int^{a(\tau)}_{a(0)} ({1\over \kappa}){a^2da\over \sqrt{a^6-r^2}}, \nonumber 
\end{eqnarray}
yields,
\begin{eqnarray}
a(\tau) = [{P_{\beta}\over \kappa}\cosh (3\kappa \tau)]^{1/3}
\end{eqnarray}
where again we chose the wormhole neck to occur for $\tau=0$ and $a(\tau=0) =
({3\over 4|\Lambda|}P^2_{\beta})^{1/6}$ and we defined $\kappa^2\equiv {4\over 3}
|\Lambda|$ and $r\equiv P_{\beta}/\kappa = P_{\beta}\sqrt{{3\over 4|\Lambda|}}$. \\
We again discuss exactly what type of topology changing process this wormhole
instanton solution describes. Thus we note ; \\
For $\tau \rightarrow -\infty$,
\begin{eqnarray}
&a^3(\tau) \rightarrow e^{-3\kappa \tau} \rightarrow \infty, ~~~~({da\over d\tau})|_{\tau
\rightarrow -\infty} \rightarrow e^{-\kappa \tau} \rightarrow \infty, \nonumber \\
&\beta(-\infty) = 0, ~~~~({d\beta \over d\tau})|_{\tau \rightarrow -\infty} =
-iP_{\beta}{1\over a^3(-\infty)} \rightarrow 0 \nonumber
\end{eqnarray}
thus the curvature scalar there goes like $R(\tau \rightarrow -\infty)
\rightarrow 12\kappa^2 = 16 |\Lambda|$, namely $\tau \rightarrow -\infty$ spacelike hypersurface
is the 3-dim. space of constant curvature with large radius, i.e., $S^3_{\infty}$. \\
For $\tau = 0$, 
\begin{eqnarray}
&a(0) = ({3\over 4|\Lambda|}P^2_{\beta})^{1/6}, ~~~~({da\over d\tau})|_{\tau = 0} = 0, \nonumber \\
&\beta(0) = -iP_{\beta}\int^{0}_{-\infty}{d\tau \over a^3(\tau)} = const.,   
~~~({d\beta \over d\tau})|_{\tau = 0} = -iP_{\beta}{1\over a^3(0)} = -i
\sqrt{{4|\Lambda|\over 3}} \nonumber
\end{eqnarray}
thus the trace of the extrinsic curvature there is again $K(\tau = 0) = 3({a'\over a})|_{\tau
= 0} = 0$, namely this $\tau = 0$ spacelike hypersurface is a surface of vanishing
extrinsic curvature and hence is the wormhole throat. And this surface has
large, constant anisotropy and constant anisotropy changing rate.
Consequently, this wormhole solution with very large anisotropy is an instanton
configuration describing the topology change from the initial ($\tau \rightarrow -\infty$)
state with topology $S^3_{\infty}$ to the final ($\tau = 0$) state with the topology 
$S^3_{\infty} \oplus S^3$, i.e., large 3-sphere with an additional closed baby universe. \\
Having constructed wormhole instanton solutions, we now turn to their contribution
to the topology-changing tunnelling amplitude. Namely, we evaluate the wormhole
instanton action, $I(instanton)$, then the quantity, $\exp{[-I(instanton)]}$, would
represent the semi-classical approximation to the topology-changing tunnelling
amplitude. Thus by substituting the ($\tau \tau$-component of) Einstein equation
in (10) satisfied by the wormhole solution into the Euclidean action in eq.(1) or (9),
one gets
\begin{eqnarray}
I(instanton) = \int^{a(0)}_{a(-\infty)} {[V(\beta) - {4\over 3}|\Lambda|a^2]a^3da
\over \sqrt{{4\over3}|\Lambda|a^6-V(\beta)a^4-P^2_{\beta}}}
\end{eqnarray}
where we used $(d\beta/d\tau) = -i(N/a^3)P_{\beta}$, 
$d\tau = a^2da/N\sqrt{{4\over3}|\Lambda|a^6-V(\beta)a^4-P^2_{\beta}}$ and the
Gibbons-Hawking surface term vanishes on each boundary, one at $\tau =-\infty$
and the other at $\tau = 0$ as we have noted earlier.
Note, first, that above expression for $I(instanton)$ is independent of the 
lapse function $N$ and hence the ``physically observable'' topology-changing
tunnelling amplitude $\exp{[-I(instanton)]}$ possesses manifest gauge-invariance
associated with the time reparametrization invariance as it should. Unfortunately,
it does not seem possible to carry out the integral in eq.(23) to obtain the 
precise value of the wormhole instanton action. Still, however, we can evaluate
$I(instanton)$ for wormhole solution with very small/large anisotropy we
constructed earlier. Firstly, for the wormhole with very small anisotropy
given in  eq.(21),
\begin{eqnarray}
I(instanton) = -\int^{a(0)}_{\lambda_{1}}{a^3da\over \sqrt{a^4-P^2_{\beta}}}
= {1\over 2}\sqrt{\lambda^4_{1} - P^2_{\beta}}
\end{eqnarray}
while secondly, for the wormhole with very large anisotropy given in eq.(22),
\begin{eqnarray}
I(instanton) = -\kappa \int^{a(0)}_{\lambda_{2}}{a^5da\over 
\sqrt{a^6-r^2}} = {\kappa\over 3}\sqrt{\lambda^6_{2} - r^2}
= {2\sqrt{3}\over 9}\sqrt{|\Lambda|\lambda^6_{2} - {3P^2_{\beta}\over 4}}
\end{eqnarray}
where $\kappa$ and $r$ are as defined earlier and $\lambda_{1}$ and 
$\lambda_{2}$ denote large cut-offs for asymptotic wormhole sizes $a(-\infty)$.
In particular, the expression for the wormhole instanton action in eq.(25) indicates 
that the tunnelling amplitude for the topology change driven by the wormhole
solution in eq.(22) gets maximized for the smallest possible value of the
cosmological constant, $|\Lambda| = 3P^2_{\beta}/4\lambda^6_{2}$. \\
We now turn to the discussion of anisotropic wormholes in the quantum regime.
The formulation we shall employ for quantum treatment of wormholes [8] can be
summarized as follows : we construct and study a minisuperspace model (again
based on the Taub cosmology) of canonical quantum cosmology in which the
main objective is to solve the Wheeler-DeWitt  equation to find the 
universe wave function. Then to see if there is an excitation that can be
interpreted as a ``quantum wormhole'', we look for a particular solution to
the WD equation with ``wormhole boundary condition'' that allows one to
identify the universe wave function as representing an excitation corresponding
to a wormhole state. And as a proposal for such wormhole boundary condition, we
shall employ the one advocated by Hawking and Page [7]. According to them, 
wormhole boundary conditions can generally be classified into two categories ;
one for the ``ground state'' and the other for the ``excited states'' of quantum
wormholes. Firstly, on the boundary condition for the ground state. When
$\sqrt{h}\rightarrow 0$, the universe wave function should be regular to represent
a non-singular 4-metric. And when $\sqrt{h}\rightarrow \infty$, the universe wave
function should be damped, say, exponentially to represent an asymptotically
Euclidean 4-metric, namely to represent the fact that there are no gravitational
excitations asymptotically. Next, on the boundary condition for the excited states.
When $\sqrt{h}\rightarrow 0$, again the universe wave function should be regular
but it may oscillates for small-$\sqrt{h}$. And when $\sqrt{h}\rightarrow \infty$,
the universe wave function should be damped.
Thus now by applying the ``Dirac quantization procedure''
to general relativity, which is one of the most well-known constraint systems, the 
classical Hamiltonian constraint given in eq.(8), turns into its quantum version,
namely the WD equation
\begin{eqnarray}
{1\over 2}[a^2{\partial^2\over \partial a^2} + (p+1)a{\partial \over \partial a}
- {\partial^2\over \partial \beta^2} + a^4\{V(\beta) + {4\over 3}\Lambda a^2\}]
\Psi[a,\beta] = 0
\end{eqnarray}
where the suffix ``$p$'' represents the well-known ambiguity in ``operator-
ordering'' and in passing from the classical Hamiltonian constraint to this
quantum WD equation, we substituted the conjugate momenta $P_{a}=-a\dot{a}/N$,
$P_{\beta}=a^3\dot{\beta}/N$ by quantum momentum operators $\hat{P_{a}}=-i
\partial/\partial a$, $\hat{P_{\beta}}=-i\partial/\partial \beta$. And of
course $\Psi[a,\beta]$ is the (physical) universe wave function. Again, we
do not intend to challenge the general solution to this WD equation. Instead,
we will be content with the universe wave function for quantum wormholes which
have specific characters we considered earlier in the classical treatment.
Namely, wormholes with very small/large anisotropy and with constant momentum
associated with the anisotropy change described by conditions given in eqs.(13)
(14) and (15). In this particular case, since $V(\beta)$ is replaced by a constant
$k$ (which is either $-1$ or $0$), the WD equation above admits a separation
of variable. Namely by setting $\Psi[a,\beta] = A(a)B(\beta)$, the WD equation
decomposes into two sectors
\begin{eqnarray}
[{d^2\over da^2} &+& {(p+1)\over a}{d\over da} + a^2(k+{4\over 3}\Lambda a^2) + 
{P^2_{\beta}\over a^2}] A(a) = 0, \\
{d^2B\over d\beta^2} &+& P^2_{\beta}B = 0
\end{eqnarray}
where we chose the constant involved in the separation of variable to be the
conjugate momentum $P_{\beta}$ (which is a constant, too) since, as is manifest
in eq.(28), it is the eigenvalue of the quantum momentum operator $\hat{P_{\beta}}
=-i\partial/\partial \beta$ which has to be the classical momentum $P_{\beta}$.
Obviously, then, the solution of $\beta$-sector is $B(\beta)=(const.)e^{iP_{\beta}
\beta}$. Next, in order to have some insight into the behavior of the solution
of $a$-sector, we view the $a$-sector of the WD equation given in eq.(27) as a
Schr\H odinger-type equation with zero total energy. Then the ``potential''
energy can be identified with
\begin{eqnarray}
U(a) = -a^2(k - {4\over 3}|\Lambda|a^2) - {P^2_{\beta}\over a^2}
\end{eqnarray}
which involves the term $(-P^2_{\beta}/a^2)$ representing an ``abyss'' in the
small-$a$ region. Since the total energy is zero, the emergence of this potential
abyss reveals the fact that (part of) the universe wave function, $A(a)$ should
be a highly oscillating function of $a$ in the small-$a$ region. And this
enormous oscillatory behavior for small scale factor $a$ appears to signal the 
existence of ``quantum wormholes'' in the small-$a$ region of the minisuperspace [8]
and hence seems consistent with the existence of classical wormhole solutions
we studied earlier. Thus in order to confirm this belief of ours, we now solve
the $a$-sector of WD equation in eq.(27) explicitly. \\
(i) Quantum wormhole with very small anisotropy ($\beta\rightarrow 0$) \\
This case amounts to choosing $k=-1$ and then an exact solution to the $a$-sector
of WD equation in (27) is available in the absence of the cosmological constant.
\begin{eqnarray}
[{d^2\over da^2} &+& {(p+1)\over a}{d\over da} - a^2 + {P^2_{\beta}\over 
a^2}] A(a) = 0, \\
A(a) &=& a^{-{1\over 2}p}Z_{i{1\over 4}\sqrt{4P^2_{\beta}-p^2}}({i\over 2}a^2),
\qquad (-2P_{\beta}<p<2P_{\beta}) \nonumber
\end{eqnarray}
where $Z_{\nu}(x)$ denotes a Bessel function of order $\nu = i{1\over 4}
\sqrt{4P^2_{\beta} - p^2}$. For now $Z_{\nu}(x)$
could be the Bessel function of the 1st kind or the 2nd kind (i.e., the Neumann
function) or their linear combinations, i.e., the Hankel functions. However,
imposing the appropriate ``wormhole boundary condition'' of Hawking and Page,
it should be, for small 3-geometry (i.e., $a\rightarrow 0$), the Bessel function
of the 1st kind since 
$A(a) = a^{-{1\over 2}p}J_{i{1\over 4}\sqrt{4P^2_{\beta}-p^2}}({i\over 2}a^2)
\rightarrow a^{-{1\over 2}p}a^{2\nu} = a^{-{1\over 2}p}\exp{[i2|\nu|\ln a]}$
(where we used $J_{\nu}(x)\rightarrow x^{\nu}/2^{\nu}\nu !$ for $x\rightarrow 0$)
with $-2P_{\beta}<p\leq 0$ for the regularity at $a=0$, which indeed possesses
highly oscillatory behavior for $a\rightarrow 0$. Next, for large 3-geometry
(i.e., $a\rightarrow \infty$), it should be the Hankel function since
$A(a) = a^{-{1\over 2}p}H^{(1)}_{i{1\over 4}\sqrt{4P^2_{\beta}-p^2}}({i\over 2}a^2)
\rightarrow a^{-{1\over 2}(p+2)}e^{-{1\over 2}a^2}$ (where we used $H^{(1)}_{\nu}
(x) = J_{\nu}(x) + iN_{\nu}(x) \rightarrow x^{-{1\over 2}}e^{ix}$ for 
$x\rightarrow \infty$), which indeed possesses rapidly damping behavior for
$a\rightarrow \infty$ as desired. Thus this solution $\Psi[a,\beta] = A(a)
e^{iP_{\beta}\beta}$ is a legitimate universe wave function of a quantum wormhole.\\
(ii) Quantum wormhole with very large anisotropy ($\beta\rightarrow \infty$) \\
This case amounts to choosing $k=0$ and an exact solution to eq.(27) is available
even in the presence of the cosmological constant. 
\begin{eqnarray}
[{d^2\over da^2} &+& {(p+1)\over a}{d\over da} - {4\over 3}|\Lambda|a^4 + 
{P^2_{\beta}\over a^2}]
A(a) = 0, \\
A(a) &=& a^{-{1\over 2}p}Z_{i{1\over 6}\sqrt{4P^2_{\beta}-p^2}}(i{2\over 
3}\sqrt{{|\Lambda|\over 3}}a^3),
\qquad (-2P_{\beta}<p<2P_{\beta}). \nonumber
\end{eqnarray}
Again, imposing the ``wormhole boundary condition'' of Hawking and Page, this
general solution of the Bessel equation should be, for small 3-geometry
($a\rightarrow 0$), the Bessel function of the 1st kind since 
$A(a) = a^{-{1\over 2}p}J_{i{1\over 6}\sqrt{4P^2_{\beta}-p^2}}
(i{2\over 3}\sqrt{{|\Lambda|\over 3}}a^3)
\rightarrow a^{-{1\over 2}p}a^{3\nu} = a^{-{1\over 2}p}\exp{[i3|\nu|\ln a]}$
with $\nu = i{1\over 6}\sqrt{4P^2_{\beta} - p^2}$ now and
$-2P_{\beta}<p\leq 0$ for the regularity at $a = 0$, which exhibits
highly oscillatory behavior for $a\rightarrow 0$. Next, for large 3-geometry
(i.e., $a\rightarrow \infty$), it should be the Hankel function since
$A(a) = a^{-{1\over 2}p}H^{(1)}_{i{1\over 6}\sqrt{4P^2_{\beta}-p^2}}
(i{2\over 3}\sqrt{{|\Lambda|\over 3}}a^3)
\rightarrow a^{-{1\over 2}(p+3)}e^{-{2\over 3}\sqrt{{|\Lambda|\over 3}}a^3}$,
which exhibits rapidly damping behavior for $a\rightarrow \infty$ as desired.
And this solution $\Psi[a,\beta] = A(a)e^{iP_{\beta}\beta}$ can be identified 
with a legitimate quantum wormhole wave function as well. Finally, we note that
the wormhole wave functions given above in (i) and (ii) can be identified with
the ones representing ``excited states'' of quantum wormholes we stated earlier
as they are regular at $a = 0$ and possess oscillatory behaviors for small-$a$. \\
To summarize, in this letter, we demonstrated explicitly that Euclidean,
anisotropic wormhole solutions do exist even in pure Einstein gravity system.
This, among others, strongly supports our intuitive expectation that there
should be generic quantum fluctuations in spacetime alone even in the complete
absence of matter. Noticing that homogeneous, isotropic wormhole solutions do
not exist in pure general relativity, one may naturally suspect that the
anisotropy degree of freedom could play a decisive role in switching on the
wormhole configuration. This is indeed the case as one can see in this work
that it is the nonvanishing $P^2_{\beta}$, i.e., the momentum or the energy
associated with the anisotropy change, that essentially renders the occurrence
of both classical and quantum wormholes possible. After all, wormholes seem
to be one of the rare means which are under our control (to some extent) to
challenge the quantum gravity.

\vspace{2cm}

{\bf \large References}

\begin{description}

\item {[1]} E. Baum, Phys. Lett. {\bf B133}, 185 (1983) ; 
            S. W. Hawking, Phys. Lett. {\bf B134}, 403 (1984).
\item {[2]} S. Coleman, Nucl. Phys. {\bf B307}, 864 (1988) ;
            Nucl. Phys. {\bf B310}, 643 (1988) ;
            I. Klebanov, L. Susskind and T. Banks, Nucl. Phys. {\bf B317},
            665 (1989).
\item {[3]} S. Giddings and A. Strominger, Nucl. Phys. {\bf B307}, 854 (1988).
\item {[4]} S. Giddings and A. Strominger, Nucl. Phys. {\bf B306}, 890 (1988).
\item {[5]} K. Lee, Phys. Rev. Lett. {\bf 61}, 263 (1988).
\item {[6]} A. Hosoya and W. Ogura, Phys. Lett. {\bf B22}, 117 (1989) ;
            S. -J. Rey, Nucl. Phys. {\bf B336}, 146 (1990).
\item {[7]} S. W. Hawking and D. N. Page, Phys. Rev. {\bf D42}, 2655 (1990).
\item {[8]} H. Kim, Nucl. Phys. {\bf B527}, 311 (1998) ; {\it ibid} {\bf B527},
            342 (1998).
\item {[9]} G. W. Gibbons and S. W. Hawking, Phys. Rev. {\bf D15}, 2752 (1977).
\item {[10]} A. H. Taub, Ann. Math. {\bf 53}, 472 (1951).

\end{description}

\end{document}